\documentclass[twocolumn]{openjournal}
\usepackage{graphicx}
\usepackage{amsmath}
\usepackage{amssymb}
\usepackage{xspace}
\usepackage[dvipsnames]{xcolor}
\usepackage[backref, pagebackref=false, hyperindex=true, breaklinks=true, colorlinks=true, urlcolor=magenta, linkcolor=NavyBlue, citecolor=cyan, citebordercolor={0 1 0}, pagecolor=red, bookmarks=true, filecolor=blue,bookmarksopen=true, plainpages=false, pdfpagemode=UseThumbs]{hyperref}
\usepackage{multirow}
\usepackage{soul}

\newcommand{\myeqref}[1]{Eq.\,(\ref{#1})}
\newcommand{\figref}[1]{Fig.~\ref{#1}}
\newcommand{\Figref}[1]{Figure~\ref{#1}}
\newcommand{\ITdd}{\textbf{ITdd}}
\newcommand{\ITdp}{\textbf{ITdp}}
\newcommand{\ITpd}{\textbf{ITpd}}
\newcommand{\ITpp}{\textbf{ITpp}}
\newcommand{\ITp}{\textbf{ITp}}
\newcommand{\ITd}{\textbf{ITd}}

\newcommand{\oca}{Université Côte d’Azur, Observatoire de la Côte d’Azur, CNRS, laboratoire Lagrange, Nice, France}
\newcommand{\mauca}{MAUCA -- Master track in Astrophysics, Université Côte d’Azur \& Observatoire de la Côte d’Azur, Parc Valrose, 06100 Nice, France}

\begin{document}

\title{On inertial forces (indirect terms) in problems with a central body}

\shorttitle{Inertial forces in problems with a central body}
\shortauthors{Crida et al.}

\author{Aurélien Crida$^1$}
\author{Clément Baruteau$^2$}
\author{Philippine Griveaud$^{1,3}$}
\author{Elena Lega$^1$}
\author{Frédéric Masset$^4$}
\author{\\William Béthune$^{5,6}$}
\author{David Fang$^{2,7}$}
\author{Jean-François Gonzalez$^8$}
\author{Héloïse Méheut$^1$}
\author{Alessandro Morbidelli$^{1,9}$}
\author{\\Fabiola Gerosa$^1$}
\author{Dylan Kloster$^1$}
\author{Léa Marques$^{10, 11, 12}$}
\author{Alain Miniussi$^{13}$}
\author{Kate Minker$^{10, 1}$}
\author{Gabriele Pichierri$^{14, 15}$}
\author{Paul Segretain$^1$\vspace{6pt}}

\affiliation{$^1$ \oca}
\affiliation{$^2$ IRAP, Universit{\'e} de Toulouse, CNRS, Université Paul Sabatier, CNES, Toulouse, France}
\affiliation{$^3$ Max-Planck-Institut für Astronomie, Königstuhl 17, 69117 Heidelberg, Germany}
\affiliation{$^4$ Instituto de Ciencias Físicas, Universidad Nacional Autónoma de México, Cuernavaca, Morelos 62210, México}
\affiliation{$^5$ Institut für Astronomie und Astrophysik, Universität Tübingen, Auf der Morgenstelle 10, 72076 Tübingen, Germany}
\affiliation{$^6$ DAAA, ONERA, Université Paris Saclay, F-92322 Châtillon, France}
\affiliation{$^{7}$ ENS de Lyon, Département de Physique, F-69342 Lyon, France}
\affiliation{$^8$ Universite Claude Bernard Lyon 1, CRAL UMR5574, ENS de Lyon, CNRS, Villeurbanne 69622, France}
\affiliation{$^9$ Collège de France, CNRS, PSL Univ., Sorbonne Univ., Paris, 75014, France}
\affiliation{$^{10}$ \mauca}
\affiliation{$^{11}$ Leibniz-Institüt für Astrophysik Potsdam (AIP), An der Sternwarte 16, D-14482, Potsdam, Germany}
\affiliation{$^{12}$ Universität Potsdam, Institut für Physik und Astronomie, Karl-Liebknecht-Str. 24-25, 14476 Potsdam, Deutschland}
\affiliation{$^{13}$ Université Côte d’Azur, Observatoire de la Côte d’Azur, CNRS, UAR Galilée, Nice, France}
\affiliation{$^{14}$ Division of Geological and Planetary Sciences California Institute of Technology, Pasadena, CA 91125, USA}
\affiliation{$^{15}$ Dipartimento di Fisica, Università degli Studi di Milano, Via Celoria 16, I-20133 Milano, Italy}

\begin{abstract}
Gravitational systems in astrophysics often comprise a body -- the primary -- that far outweights the others, and which is taken as the centre of the reference frame.
A fictitious acceleration, also known as the indirect term, must therefore be added to all other bodies in the system to compensate for the absence of motion of the primary.
In this paper, we first stress that there is not \emph{one} indirect term but as many indirect terms as there are bodies in the system that exert a gravitational pull on the primary. 
For instance, in the case of a protoplanetary disc with two planets, there are three indirect terms: one arising from the whole disc, and one per planet.
We also highlight that the direct and indirect gravitational accelerations should be treated in a balanced way: the indirect term from one body should be applied to the other bodies in the system that feel its direct gravitational acceleration, and only to them.
We point to situations where one of those terms is usually neglected however, which may lead to spurious results.
These ideas are developed here for star-disc-planets interactions, for which we propose a recipe for the force to be applied onto a migrating planet, but they can easily be generalized to other astrophysical systems.
\end{abstract}

  
\maketitle


\section{Introduction}

To include or not to include indirect terms?
The question arises in many simulations of astrophysical problems where one object of a gravitational system far outweighs the others.
Examples include a planet and its satellites, a star and its protoplanetary disc and planets, or a (supermassive) black hole and its accretion disc.
The most massive object, which is often called the primary, largely dominates the gravity in the system, and the simplest description of the motion is to say that the other objects in the system simply orbit around the primary.
This description is, of course, approximate since only the centre-of-mass of the system has no acceleration and thus remains fixed.
Nevertheless, for simplicity, many studies write equations of motion in a frame centred on the primary, which then becomes \emph{literally} the central object in the system.
Fictitious forces known as inertial forces, and dubbed \emph{the indirect term}, are therefore added to account for the primary's motion with respect to an inertial frame, and to properly describe the dynamics of the whole system.
The concept of inertial forces should be an easy aspect of gravitational dynamics, yet the literature is rather vague, if not confusing, in situations where objects in the system have very small masses compared to the primary's.

The purpose of this manuscript
is to provide a clear, pedagogical picture of how to deal with inertial forces in gravitational systems with large mass ratios amongst their constituents, in particular in numerical simulations.
An immediate, visual example is that of the restricted three-body problem comprised of a star, a planet and massless "test" particles, for which neglecting the reflex motion of the star imparted by the planet would miss the Lagrange points L4 and L5 as equilibrium positions (see \S~\ref{sub:ITpd}).
Another example is that of a protoplanetary disc around a star where the disc-to-star mass ratio is small enough for the disc's self-gravity to be discarded for simplicity.
In that case, we argue that the inertial force due to the action of the disc onto the star\footnote{This force cancels out if the mass distribution in the disc is axisymmetric.} 
should not be applied to the disc, in order to consistently neglect the disc's self-gravity.
In our companion paper (Crida et al., in revision, hereafter Paper II), we report on a long-term physical instability which we attribute to this very inertial force in stellocentric simulations.
This instability, however, sets in whether or not disc self-gravity is included.

This paper is organized as follows.
In section~\ref{sec:physics}, we review in detail the physics of the problem and set out clearly why inertial forces are important. 
This section serves to illustrate that, in some cases, taking inertial forces into account or not is not such an easy question.
We show that the indirect term should be split into its components associated to each constituent in the system for a clear understanding and, in section~\ref{sec:appl}, we specialize to star-disc-planets interactions in protoplanetary discs. 
We study in detail the four possible cases where the contributor and the recipient of the indirect term are a planet or the disc.
In particular, we propose a recipe for the force to be applied onto a migrating planet.
May this help for the clarification and reproducibility of everyone's work.

\section{Physics of the problem}
\label{sec:physics}

\subsection{Back to basics}
\label{sub:basics}

The first principle of classical mechanics, a.k.a. Newton's first law, states that a pseudo-isolated object (which feels a total force of zero) stays at rest if it is at rest, and follows a uniform rectilinear motion with constant velocity vector if it is in motion.
This statement, however, only applies in so-called Galilean or inertial frames.
In a frame which undergoes an acceleration, a pseudo-isolated object appears to experience an opposite acceleration\footnote{For instance, when a bus brakes (backwards acceleration), passengers are projected forward in the frame of the bus, because they tend to keep moving at constant velocity in the Galilean frame of the road.}.
To describe the motion of a system in a non-inertial frame, one needs to add a fictitious acceleration which imprints to every object in the system an acceleration opposite to that of the centre of the frame\footnote{If the frame is rotating, one should also add the centrifugal and Coriolis accelerations. 
Accounting for these forces is not the subject of the present work, and is generally done, in disc simulations, in a way that conserves angular momentum \citep{Kley1998}.}.
This fictitious, inertial acceleration is often called \emph{the indirect term}\footnote{Note that some authors may prefer to call \emph{the indirect term} the potential from which this acceleration derives.}, at least in astrophysical problems with a dominant central object.

Using bold letters to denote vectors, the indirect term acceleration in a gravitational system can be written as
\begin{equation}
    \mathbf{IT} = -\mathbf{{a}_{*}} = \sum_i \underbrace{-\frac{GM_i}{r_i^3}\mathbf{r_i}}_{\mathbf{IT\emph{i}}}
    \label{eq:IT}
\end{equation}
where $\mathbf{{a}_{*}}$ is the acceleration of the non-inertial frame (that is, of the central object, corresponding to the star subscript) with respect to the centre-of-mass of the system. 
It can be decomposed as the sum of the accelerations imprinted by all objects in the system, with $M_i$ the mass of object $i$ and $\mathbf{r_i}$ its position vector.
This decomposition highlights that to each object can be associated an indirect term, which will turn out to be a key concept here.

\subsection{Tides}

\begin{figure}
    \centering
    \includegraphics[width=\linewidth]{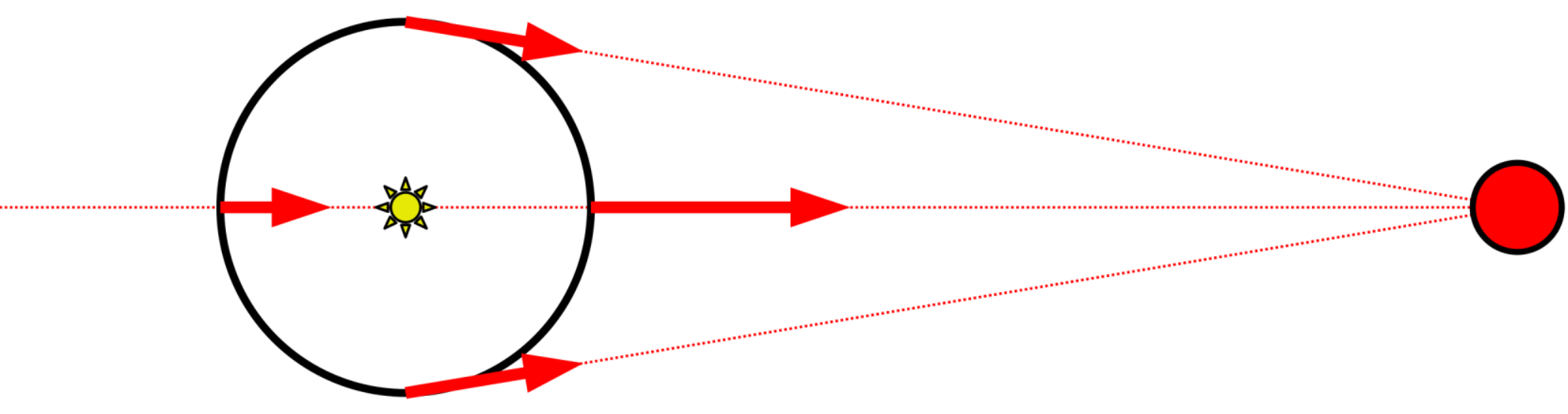}\vspace{8pt}
     \includegraphics[width=\linewidth]{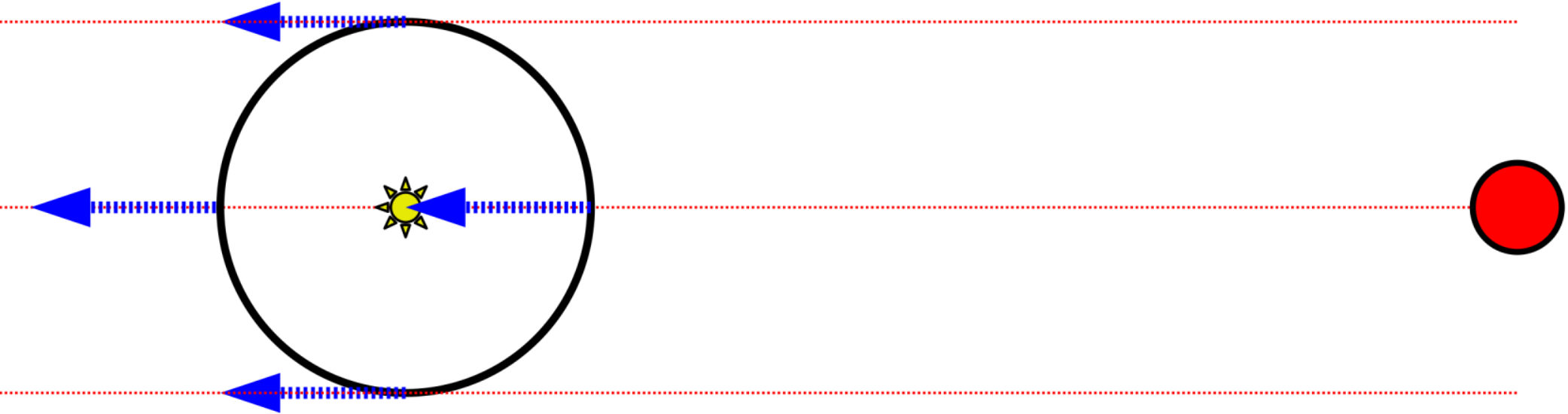}
    \includegraphics[width=\linewidth]{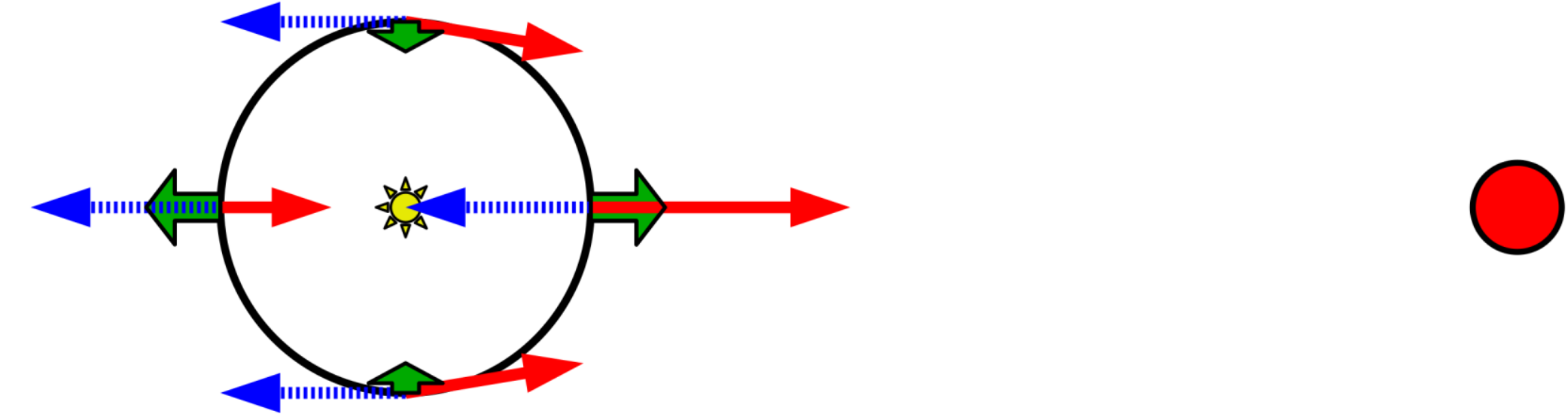}\vspace{8pt}
    \caption{Accelerations imparted by a distant companion (red disc) on a ring (black circle) around a star, in a frame centred on the star (yellow star symbol). Red arrows (top): direct gravitational acceleration of the companion. Blue dashed arrows (middle): indirect gravitational acceleration of the companion, or indirect term, that is the opposite of the acceleration exerted by the companion on the star. Green thick arrows (bottom): resulting tidal acceleration.}
    \label{fig:ring}
\end{figure}

A well-known manifestation of the indirect term is the tidal force.
Consider a narrow ring around a primary object, in the presence of a distant companion (secondary).
The companion's mass is assumed to be much smaller than the primary's, and its distance much larger than the ring's radius.
It is the situation depicted in \figref{fig:ring}, where the companion is shown as a red disc.
What is the effect of the secondary on this ring?

Let us adopt a frame of reference centred on the primary.
First, the ring feels the \emph{direct} gravitational acceleration of the secondary, which attracts the whole ring towards it (see the red arrows in the top panel of \figref{fig:ring}, which are all directed towards the secondary, and differ in length depending on the distance to the secondary).
Similarly, the star is accelerated towards the secondary; so, in our frame the opposite of the acceleration exerted by the secondary on the star applies to the ring (see blue arrows in the middle panel of \figref{fig:ring}, which are all identical in length and direction). 
It is this very acceleration that is named "indirect term", and which is an inertial, fictitious acceleration.
We could be a little more specific here and call it the "indirect term due to the secondary".
The small difference between the direct and indirect gravitational accelerations of the secondary results in a net acceleration on the ring that corresponds to the tidal acceleration\footnote{Note that, averaged over the ring, the tidal acceleration cancels out.} (see green arrows in the bottom panel of \figref{fig:ring}).
On the Earth this effect leads to the well known semi-diurnal excitation of the sea level.
We further stress that, considering only the direct (indirect) gravitational acceleration of the secondary would lead to a non-physical situation where the whole ring would be pulled in (away from) the direction of the companion instead of being slightly stretched around the star.

To wrap up this paragraph, we stress that there is no question that both direct and indirect gravitational accelerations must be considered for a proper physical description of the problem. 
The discussion above simply highlights the equally critical roles of both of them. 
Yet, in the following subsection, we will see that sometimes, in numerical simulations, only one of the two terms is actually computed.

\subsection{Hierarchical or non fully self-gravitating situation}

Special care should be taken in the computation of indirect terms in systems where there is a mass hierarchy amongst their constituents.
It is common indeed to discard the \emph{direct} gravitational acceleration due to the least massive objects because it has (very) little impact on the overall dynamics and it saves a considerable amount of computing time. 
But it raises the issue of how to deal with their \emph{indirect} counterpart.
A classical example is that of N-body simulations of planetary systems with so-called \emph{small particles}, which can be asteroids for instance: they have non-zero mass and do influence planets, but their influence on one another can be neglected.
In this way, the number of gravitational interactions to be computed increases linearly with the number of small particles and not quadratically.

Let us take an illustrative approach by considering a gravitational system comprised of a massive central object (a star), a less massive object $A$ (e.g., a planet), and two even less massive objects $B$ and $C$ (which can be gas parcels in the protoplanetary disc around the star or simply small solids like asteroids).
If $A$ feels the gravity of $B$ and $C$, but $B$ and $C$ do not feel each other's gravity, should the acceleration of the central object due to $B$ and $C$ ($\mathbf{IT_B}$ and $\mathbf{IT_C}$) be included in the indirect term?
To answer this question, let us look again at \figref{fig:ring}, assume that $C$ is the red disc and that the black circle is the orbit of $A$.
As we have seen in the previous subsection, since $A$ feels the \emph{direct} gravitational acceleration from $C$ (red arrows), it should also feel the \emph{indirect} gravitational acceleration from $C$ (blue arrows).
So $A$ should indeed feel $\mathbf{IT_C}$.
And, similarly, $A$ should feel $\mathbf{IT_B}$.
But now let us assume that the black circle is the orbit of $B$ and $C$ is the red disc again.
As we assume that $B$ and $C$ do not feel each other's gravity, $B$ does not feel the \emph{direct} gravitational acceleration from $C$ (no red arrows in \figref{fig:ring}).
If $B$ felt the \emph{indirect} gravitational acceleration from $C$ (i.e. the blue arrows in \figref{fig:ring}), its orbit would be pulled away from $C$ as a whole.
To avoid this, we see that $B$ should \emph{not} feel $\mathbf{IT_C}$.
And, similarly, $C$ should \emph{not} feel $\mathbf{IT_B}$.

So how to compute $\mathbf{IT}$ in the end?
From the above example, we see that gravity in a non-inertial frame should be treated in a balanced fashion: if the direct gravitational acceleration is taken in, so must be its associated indirect term.
If, however, the direct gravitational acceleration is discarded for simplicity, so should be its corresponding indirect term.
To be or not to be included in the indirect term, that seems to be a solved question, if one admits that there is no such thing as \emph{the} indirect term, but that different objects in the system (like $A$, $B$, and $C$ above) may feel different indirect terms.

\section{Applications}
\label{sec:appl}
In the previous section, we have stressed that, in a gravitational system, the indirect term -- the opposite of the gravitational acceleration on the central object -- can be split into contributions from all constituents in the system that gravitationally pull the central object.
We have argued that, when the direct gravitational force from one constituent in the system is discarded, so should be its corresponding indirect term.
In this section, we illustrate our proposal based on a specific gravitational system: a protoplanetary disc around a young star with one or more embedded planets.

\begin{figure}
    \centering
    \includegraphics[width=\linewidth]{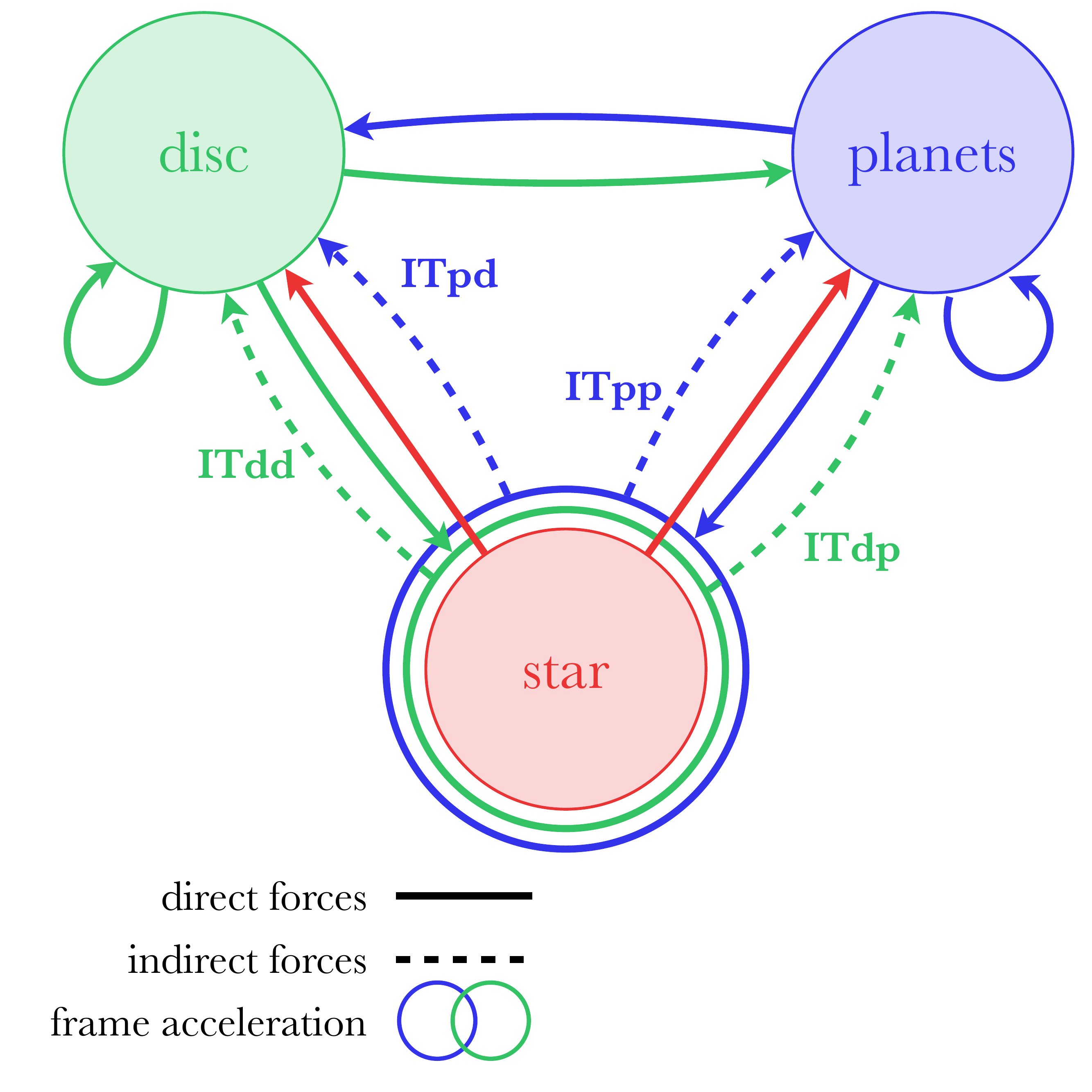}
    \caption{Summary of all forces applied in a gravitational system comprised of a star, a protoplanetary disc and a planet, in a frame centred on the star. Solid arrows show direct gravitational forces, while dashed arrows show the corresponding indirect forces. Two circles, blue and green, surround the star that symbolise the acceleration felt by the star from the gravitational pull of the corresponding objects (blue for the planet and green for the disc). This acceleration is the driver of the indirect terms felt by each object.}
    \label{fig:summary}
\end{figure}

Such a system features four indirect terms: the indirect gravitational acceleration of the planet(s) on the disc (\S~\ref{sub:ITpd}), the indirect gravitational acceleration of one planet on another (\S~\ref{sub:ITpp}), the indirect gravitational acceleration of the disc on the planet(s) (\S~\ref{sub:ITdp}), and the indirect gravitational acceleration of the disc on itself (\S~\ref{sub:ITdd}).
These four indirect terms, and their direct gravitational counterparts, are sketched in \figref{fig:summary}.
We extend the notations introduced in the previous section by denoting each indirect term as \textbf{ITab}, where the first subscript indicates which object in the system exerts the indirect gravitational acceleration, and the second subscript which one feels it.
Here the a and b subscripts are either 'p' for planet or 'd' for disc.
For instance, \ITpd\ denotes the indirect term exerted by the planet on the disc.

\subsection{Indirect term from planet to disc (\ITpd)}
\label{sub:ITpd}

\begin{figure*}
    \includegraphics[width=0.33\hsize]{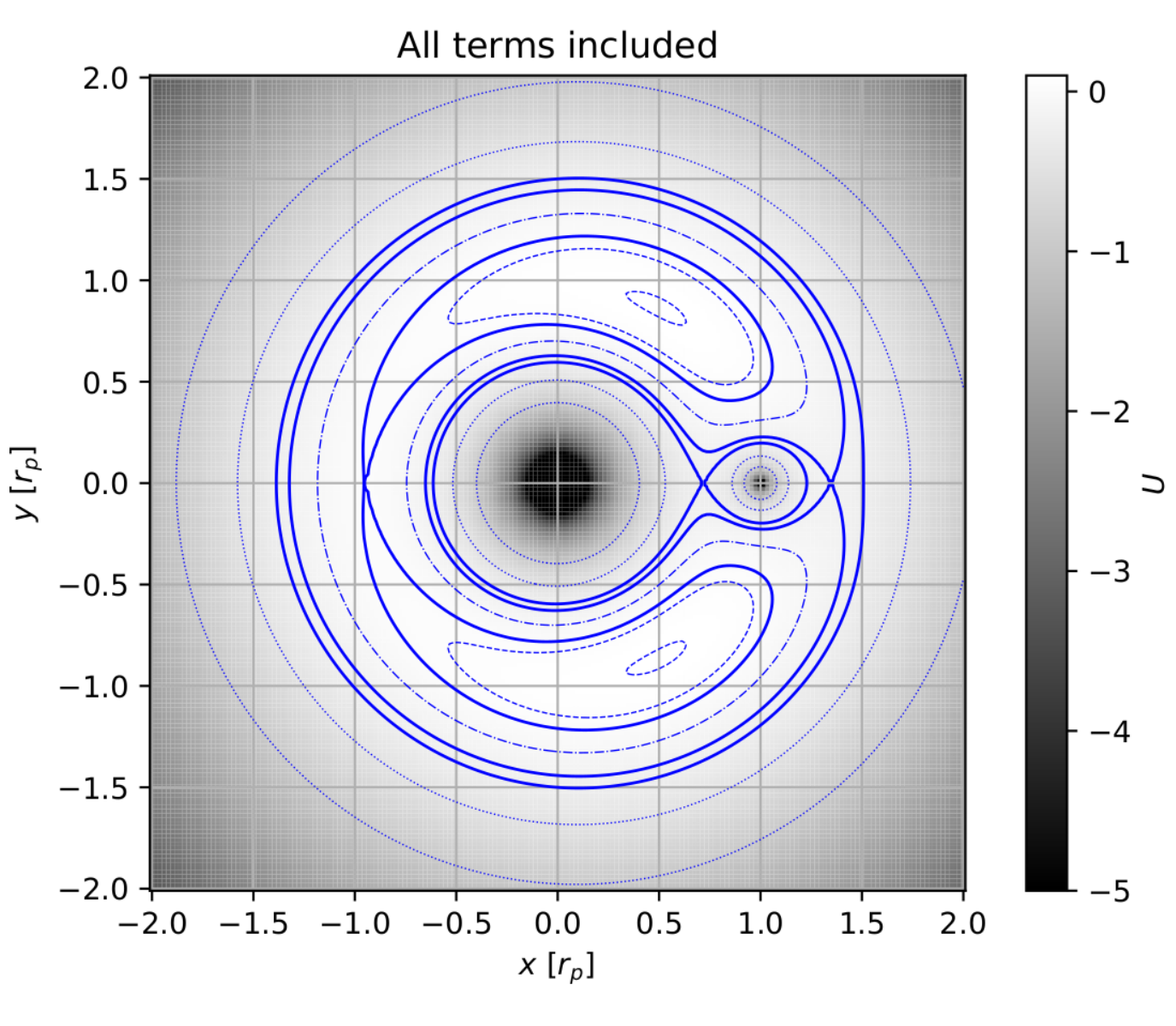}
    \includegraphics[width=0.33\hsize]{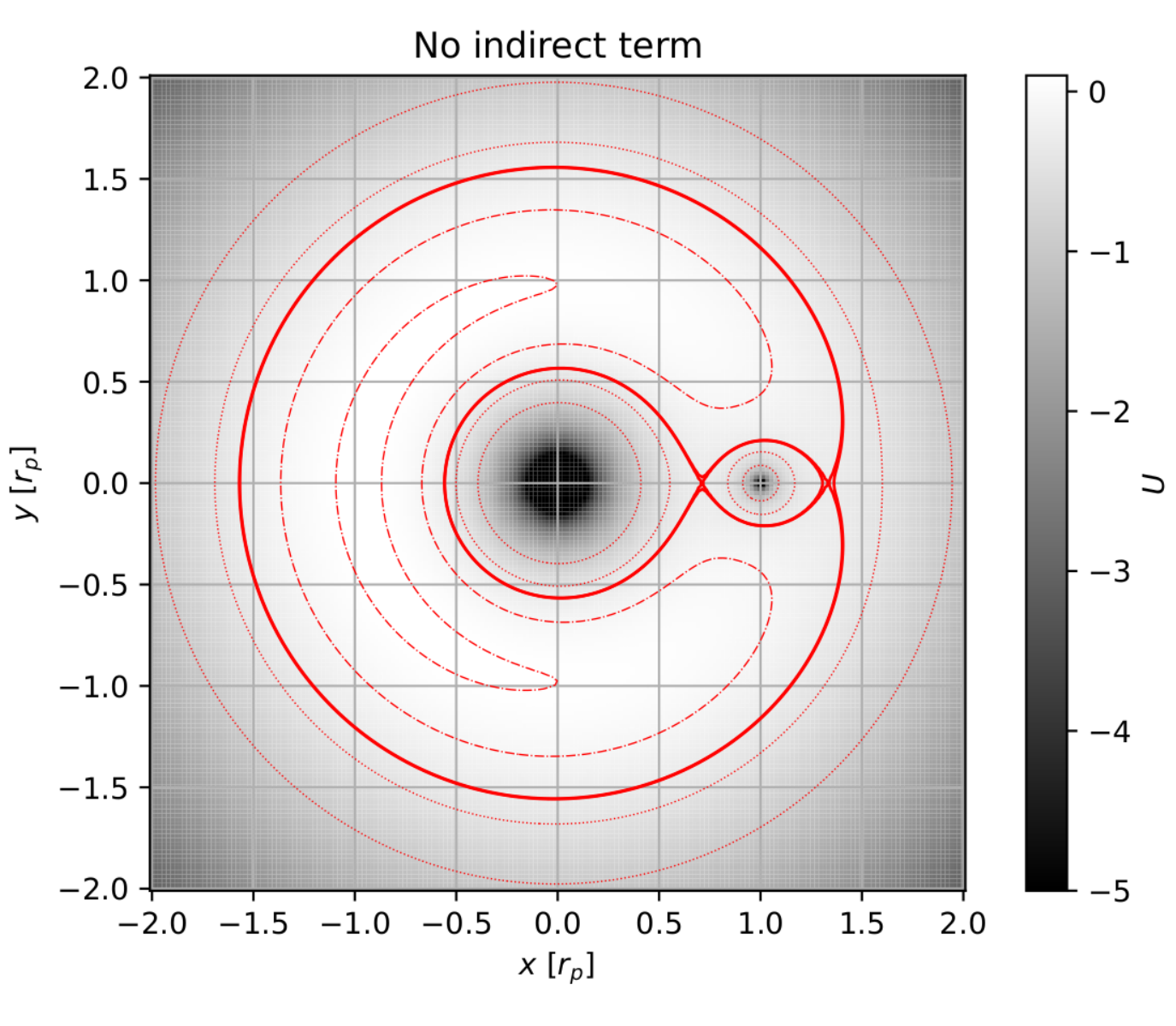}
    \includegraphics[width=0.33\hsize]{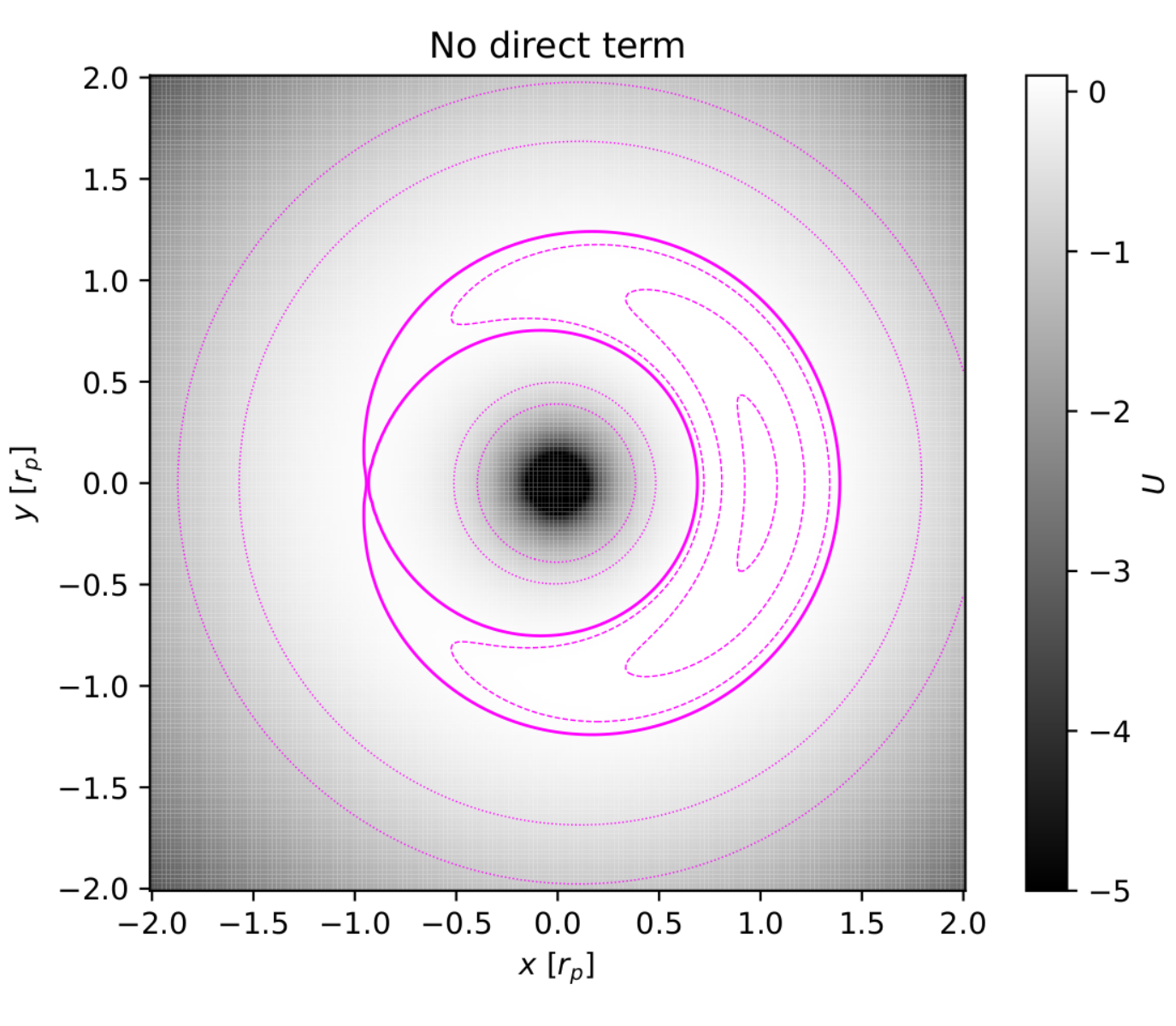}
    \caption{Zero-velocity curves (contours of potential energy) in the restricted circular three-body problem, computed in a frame centred on the star and corotating with a companion of $0.1$ stellar mass. The indirect term of the planet on the test particles is not taken into account in the middle panel, but is in the left and right panels\,; in the right panel, the direct term from the companion is removed.
    Dotted lines: circulating orbits. Dash-dotted lines: horseshoe orbits (zero-velocity curves are not actual orbits, but close enough). Solid lines: separatrices. Dashed lines: tadpole orbits.}
    \label{fig:RCTBP}
\end{figure*}

In this first subsection, we assume that the protoplanetary disc is made only of massless ``test'' particles that interact with the star and a single planet. 
The dynamics of such particles is classically described in the framework of the restricted three-body problem.
We consider here the case of a planet on a circular orbit and look at the balance of accelerations in the frame centred on the star and corotating with the planet at angular frequency $\Omega_p=\sqrt{G(M_*+M_p)/{r_p}^3}$. 
Here and in the following, $G$ denotes the gravitational constant, $M_*$ the mass of the star, $M_p$ the mass of the planet and $r_p$ the radial distance between the star and the planet. 
Note that $M_p$ in the formula comes from \ITp, see Appendix~\ref{sub:appA}.

\Figref{fig:RCTBP} displays zero-velocity curves obtained with $M_p/M_*=0.1$, a large planet-to-star mass ratio that is not quite typical of planetary systems, but which is adopted only for the sake of legibility.
Contours of potential energy (zero-velocity curves) are shown in each panel that include the gravity from the star $-GM_*/r$, the centrifugal potential energy $-\frac12 r^2{\Omega_p}^2$, the direct gravity from the planet $-GM_p/d$ (where $d$ is the distance to the planet location, here at $x_p=r_p$, $y_p=0$) except in the right panel, and the indirect term from the planet $(GM_p/{r_p}^2)x$ except in the middle panel.
Without \ITp\ (middle panel), zero-velocity curves show that there would be only 3 equilibrium locations for test particles: the Lagrange points $L_1$ to $L_3$, which are all along the star-planet direction. 
The reason for this is that, in the disc midplane, the gravitational acceleration of the star and the centrifugal acceleration are strictly parallel, being oriented in the radial direction which defines the unit vector $\mathbf{\hat{r}}$. 
Therefore, when only the direct gravitational acceleration from the planet is accounted for, there cannot be any equilibrium points away from the straight line joining the star and the planet. 
Now, the indirect term from the planet,
\begin{equation}
    \mathbf{ITp} = -\mathbf{{a}_{*,p}} 
    = - \left(\frac{GM_p}{{r_p}^2}\right) \mathbf{\hat{x}}\ ,
    \label{eq:ITp}
\end{equation}
is oriented along the star-to-planet direction defined by unit vector $\mathbf{\hat{x}}$, not $\mathbf{\hat{r}}$.
So only by including \ITp\ (left panel) can we also recover the existence of the Lagrange points $L_4$ and $L_5$, located at $\pm 60$ degrees ahead of, and behind the planet in its orbit, and which are well-known in the Solar System for hosting trojan asteroids (in particular along Jupiter' orbit). 
The right panel, in which the direct term is removed, highlights the contribution of \ITp\ to the shape of the zero-velocity curves shown in the left panel.
In fact, the sum of the centrifugal acceleration in the star-centred frame and of the indirect term equals the centrifugal acceleration in the barycentric frame.
One may then think that it would be more appropriate and easier to simply always use a barycentric frame, as most N-body codes do. However, ensuring that the primary is always at the centre of the frame has many advantages, especially for grid codes. For instance, in presence of a distant companion, the star-barycentre distance may be larger than the radius of the grid's inner edge, so that the star ends up being in the grid!

To conclude this subsection, it is clear that in order to study properly the disc response to the perturbation of a planet in a frame centred on the star, one must include \ITpd.
This is almost always done in the literature.

\subsection{Indirect term from one planet to another (\ITpp): indirect capture in mean-motion resonance}
\label{sub:ITpp}
In this second example case, we consider a system comprised of a star and of two planets, of masses $M_1$ and $M_2$, and position vectors $\mathbf{r_1}$ and $\mathbf{r_2}$ with respect to the star. 
From the previous subsection and Appendix~\ref{sub:appA}, it is clear that, to model their gravitational interactions properly in a star-centred frame, the planets should feel their own indirect term as well as each other's indirect term. 
Now, there are situations in which one may purposely deactivate the direct gravitational interaction between the planets, so that they do not feel each other's gravity. This has been done, in particular, to assess the impact of the resonant interaction between two planets by comparing their paths to those obtained in a test calculation in which their mutual gravitational interaction is suppressed, thereby shutting off, in principle, the resonance \citep[e.g.][]{BP13}.
In this case, what indirect terms should be included?

\begin{figure*}
    \includegraphics[width=\hsize]{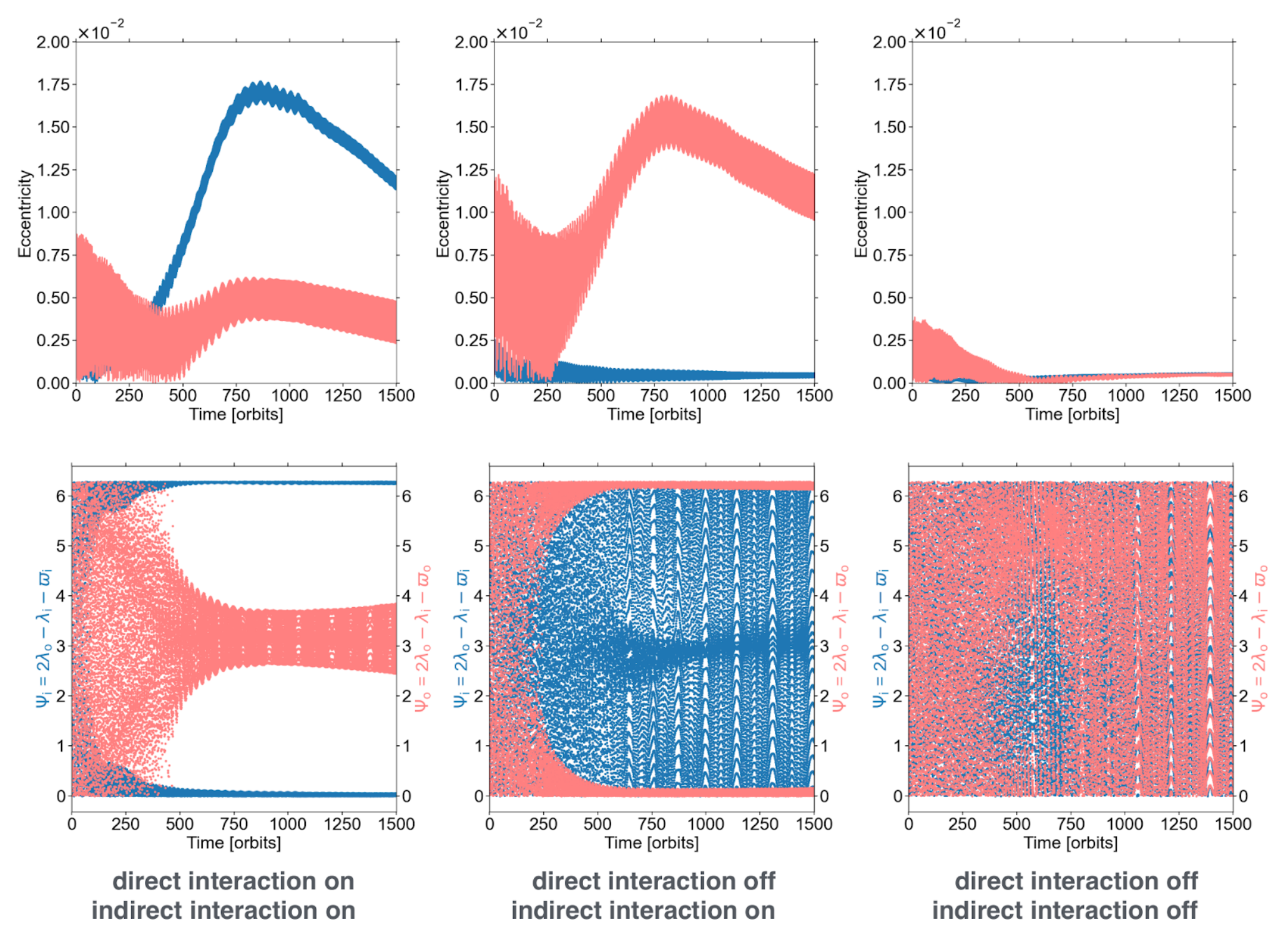}
    \caption{Time evolution of the eccentricity (top) and resonant angles of the 2:1 mean-motion resonance (bottom) for two massive planets undergoing convergent migration in their protoplanetary disc (see text), with or without including the direct and/or indirect gravitational accelerations between the planets (see labels below each column of panels). In the upper panels, blue curves are for the inner planet, red curves for the outer planet. In the lower panels, the resonant angle which involves the longitude of pericentre of the inner (outer) planet is displayed in blue (red). Part of the results shown is adapted from \citet{BP13}.}
    \label{fig:bp13}
\end{figure*}

Denoting $\mathbf{d} = \mathbf{r_2} - \mathbf{r_1}$, the \emph{direct} gravitational acceleration exerted on planet 2 due to the star and planet 1 is $\mathbf{a_{2,\rm direct}}=-GM_*\mathbf{r_2} / {r_2}^3 - GM_1\mathbf{d} / {d}^3$.
Planet 2 also experiences an \emph{indirect} gravitational acceleration from itself and from planet 1, which corresponds to the opposite of the total acceleration imprinted on the central star: $\mathbf{a_{2,\rm indirect}} = -GM_1\mathbf{r_1}/{r_1}^3 - GM_2\mathbf{r_2}/{r_2}^3$.
The total acceleration felt by planet 2 thus reads:
 \begin{align}
     \mathbf{a_2} = &-\left(\frac{G(M_*+M_2)}{{r_2}^3} + \frac{GM_1}{d^3}\right)\mathbf{r_2} \nonumber \\
     &-GM_1\left(\frac{1}{{r_1}^{3}}-\frac{1}{d^3}\right) \mathbf{r_1}.
     \label{eq:a_2}
 \end{align}
Since it is oriented along $\mathbf{r_2}$, the first term on the right-hand side of \myeqref{eq:a_2} does not torque planet 2, meaning that it does not change planet 2's angular momentum. 
What it does is to set planet 2's angular velocity to a value slightly larger than that of a purely circular motion with the star.
Only the second term on the right-hand side of \myeqref{eq:a_2} -- that in $\mathbf{r_1}$ -- can torque planet 2 and change its angular momentum.
If the direct gravitational interaction of planet 1 on planet 2 is swiched off, the term $GM_1 \mathbf{r_1} / d^3$ will cancel out. 
Now if the indirect term of planet 1 is \emph{not} swiched off, the term $-GM_1 \mathbf{r_1} / r_1^3$ will change planet 2's angular momentum.
Similarly, if the direct gravitational interaction of planet 2 on planet 1 is swiched off, but not planet 2's indirect term, the latter will still be able to change planet 1's angular momentum!
In fact, with only the indirect term from a planet, the other one feels the potential shown in the right panel of fig.~\ref{fig:RCTBP}, which is not axisymmetric.

The most spectacular, or at least unexpected consequence of this is that the planets can get locked in mean-motion resonance despite their \emph{direct} gravitational interaction being switched off!
This situation is illustrated in \figref{fig:bp13}, which shows the time evolution of the eccentricity (upper panels) and of the 2:1 resonant angles (lower panels) for two planets migrating in their protoplanetary disc: a 0.6 Jupiter-mass inner planet, and an 0.4 Jupiter-mass outer planet.
The simulations use the same disc's physical model and numerical setup as in \citet{BP13}. 
As the outer planet migrates inwards in the disc faster than the inner planet, the two planets orbital period ratio decreases and approaches $2$.
When both the direct and indirect accelerations are taken into account (left panels), the planets become locked in 2:1 resonance, as evidenced by the libration of both resonant angles and by the increase in the eccentricities (the final decrease is due to wake-planet interactions, as explained in \citealp{BP13}).
When the direct acceleration between the planets is off, but the indirect terms are still on, the outer planet is in resonance with the reflex motion of the central star around the centre-of-mass of the system imprinted by the inner planet. 
This explains the increase in the outer planet's eccentricity and the libration of the critical angle featuring the longitude of pericentre of the outer planet (middle panels).

Finally, only by further discarding the planets indirect term are the planets not able to lock in resonance (right panels). 
We finally mention that indirect terms can only lock planets in 2:1 resonance, but not in other first-order resonances like the 3:2, as we have checked numerically. Analytic explanations on this can be found in Appendix~\ref{sub:appB}. Higher order (N:1) resonances can also be excited by the indirect term, as shown in Appendix~B of \citet{HaddenTamayo2022}.

\subsection{Indirect term from disc to planet (\texttt{ITdp}):\\indirect torque}
\label{sub:ITdp}
In this section, we come back to a system composed of a star, a single planet, and its protoplanetary disc, in a reference frame centred on the star.
We denote by $\Sigma$ the disc's surface density, by $\mathbf{r}$ the position vector of a disc element, and by $\mathbf{r_p}$ the position vector of the planet.
The disc exerts on the planet a \emph{direct} gravitational acceleration that reads
\begin{equation}
\mathbf{a_{\rm dir}} = \int_{\rm disc} \frac{G \Sigma(\mathbf{r}) (\mathbf{r} - \mathbf{r_p})}{ \lVert \mathbf{r} - \mathbf{r_p} \rVert^3 } {\rm d}^2\mathbf{r},
\label{eq:acc_direct}
\end{equation}
which is associated with a \emph{direct} specific torque on the planet:
\begin{equation}
\Gamma_{\rm dir} = \left( \mathbf{r_p} \wedge \mathbf{a_{\rm dir}} \right) \cdot \mathbf{\hat{z}},
\label{eq:Gamma_direct}
\end{equation}
where $\mathbf{\hat{z}}$ is the unit vector normal to the disc plane.
As argued above, if the planet feels the direct gravitational acceleration of the disc, it should also feel the indirect acceleration from the disc:
\begin{equation}
    \mathbf{ITd} = - \int_{\rm disc} \frac{G \Sigma(\mathbf{r}) \mathbf{r}}{ \lVert \mathbf{r} \Vert^3} {\rm d}^2\mathbf{r},
    \label{eq:ITd}
\end{equation}
which is also associated with an \emph{indirect} specific torque on the planet:
\begin{equation}
\Gamma_{\rm ind} = \left( \mathbf{r_p} \wedge \mathbf{ITd} \right ) \cdot \mathbf{\hat{z}}.
\label{eq:Gamma_indirect}
\end{equation}
In simulations of disc-planet interactions, it is common to hold the planet on a fixed orbit and infer its orbital evolution by computing the torque exerted by the disc.
This is classically done by solely measuring $\Gamma_{\rm dir}$, given by \myeqref{eq:Gamma_direct}.
But when a planet moves in its disc, it feels both the direct and indirect accelerations of the disc, so the relevant torque that changes the planet's orbit is actually $\Gamma_{\rm dir} + \Gamma_{\rm ind}$.
The planet thus migrates at a rate that differs from that expected from $\Gamma_{\rm dir}$ only. 
The next three paragraphs aim at estimating the indirect torque $\Gamma_{\rm ind}$ in two regimes of planet migration, before concluding on how the disc acceleration on the planet should be computed in non self-gravitating disc simulations with a migrating planet.

\begin{figure}
    \centering
    \includegraphics[width=\hsize]{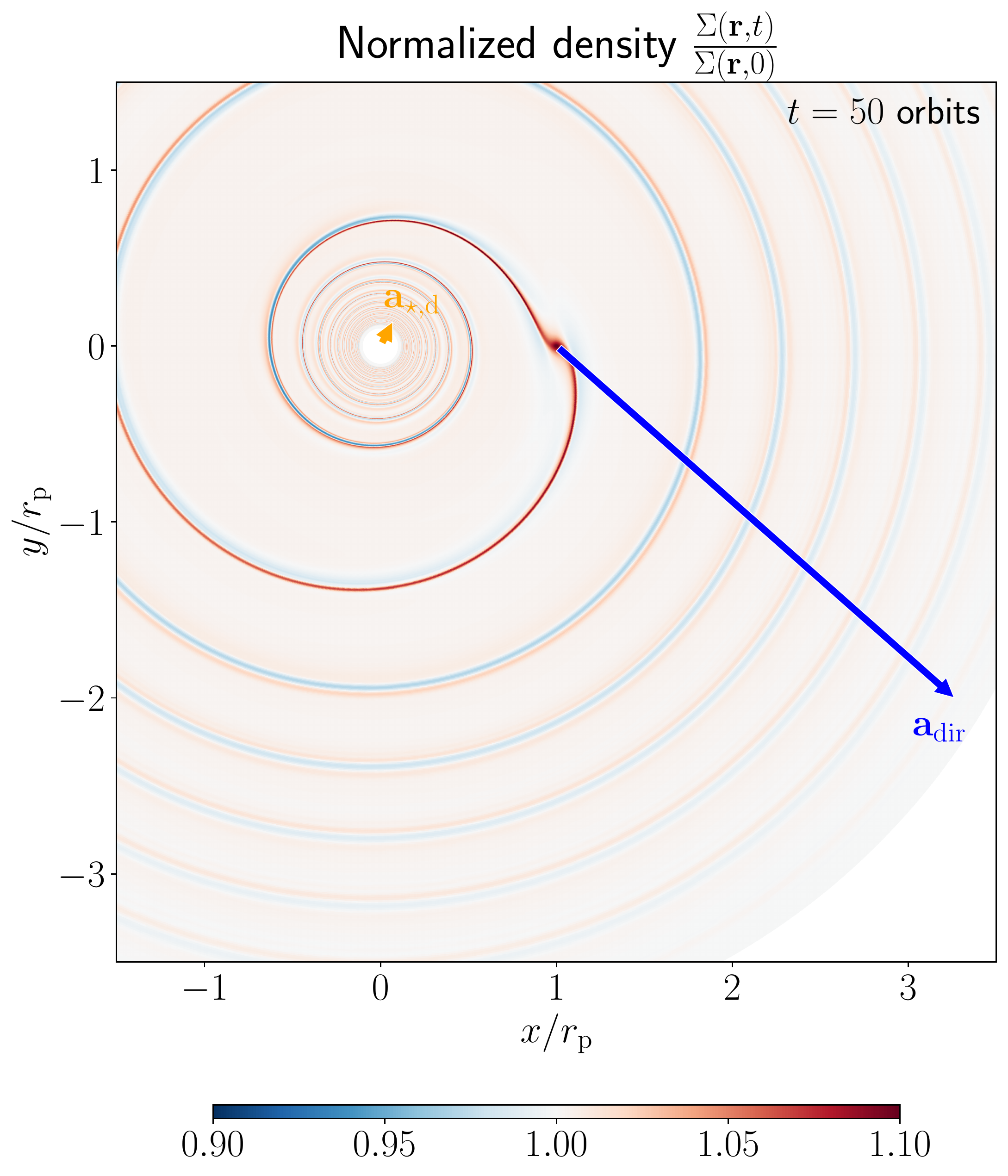}
    \includegraphics[width=\hsize]{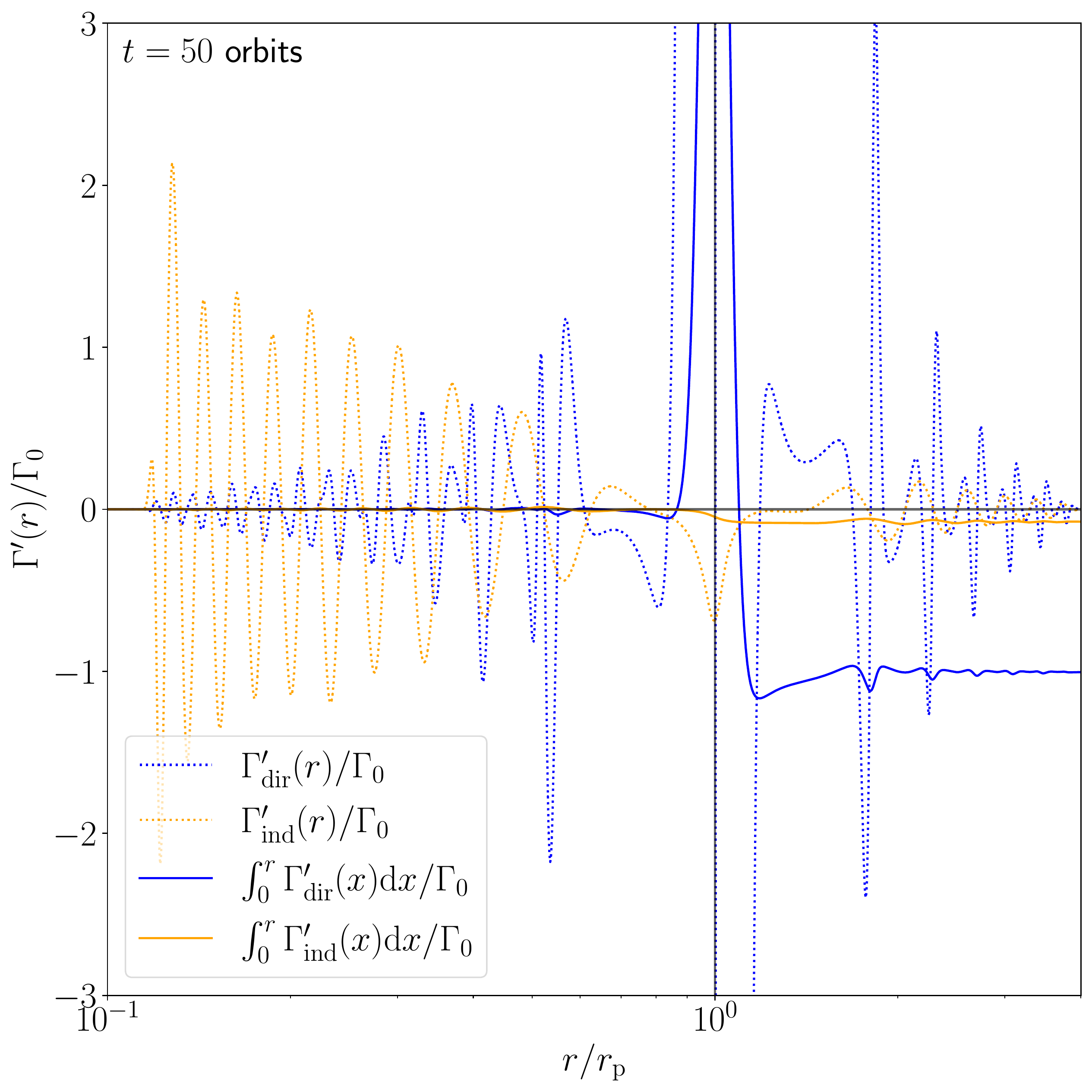}
    \caption{Top: surface density of the disc gas normalised by its initial density in a simulation of a low-mass planet (see text). Arrows show the acceleration of the star due to the disc $\mathbf{a_{*,d}}$, and the acceleration of the planet due to the direct gravity of the disk $\mathbf{a_{dir}}$ (non-axisymmetric component only, that is the first term of \myeqref{eq:a_p_disc}\,). The scale of the arrows is such that an acceleration of $2\times10^{-7}$ code units corresponds to an arrow of length $1$. Bottom: analysis of the direct and indirect torques exerted by the disc on the planet (blue and orange, respectively); dotted curves: radial torque density; solid curves: integrated torque densities.}
    \label{fig:IT5}
\end{figure}

\paragraph{Type~I migration in a smooth disc:\\link between indirect torque and corotation torque}
\,\null\ \\We consider a low-mass planet relevant to the so-called type I migration regime (for a review on planet migration, see for instance \citealp{Baruteau+PPVI}).
In this regime, the torque felt by the planet should be proportional to a standard reference torque $\Gamma_0$ (see Eq.~2 in \citealp{Baruteau+PPVI}).
We carried out a 2D hydrodynamical simulation with the public code FARGO \citep{FARGO}, whose physical and numerical parameters are given in table~\ref{tab:simu} (for simplicity an isothermal equation of state was adopted, whereby the gas temperature stayed uniform and stationary).
\begin{table}
    \centering
    \begin{tabular}{|r||c|c|}
    \hline
    PARAMETER & TYPE I & TYPE II\\
    \hline
    \hline
    Planet to star mass ratio & $q=10^{-5}$ & $q=10^{-3}$\\
    Disc aspect ratio & \multicolumn{2}{c|}{$h(r)=0.05 \times (r/r_{\rm p})^{1/2}$}\\
    Turbulent viscosity parameter & \multicolumn{2}{c|}{$\alpha = 10^{-3}$}\\
    Initial surface density profile & \multicolumn{2}{c|}{$\Sigma(r) \propto r^{-1/2}$}\\
    $\mathbf{ITdd}$ & \multicolumn{2}{c|}{discarded}\\
    Planetary orbit & \multicolumn{2}{c|}{$r_{\rm p}=1$, fixed}\\
    \hline
    Grid's radial extent & $0.1<r<4$ & $0.2<r<4$\\
    Resolution & \multicolumn{2}{c|}{$dr/r = d\theta = 3\times 10^{-3}$}\\
    Radial boundary condition & \multicolumn{2}{c|}{wave damping}\\
    \hline
    \end{tabular}
    \caption{Parameters for the simulations in \S\,\ref{sub:ITdp}.}
    \label{tab:simu}
\end{table}
\Figref{fig:IT5} shows the density perturbation obtained after 50 planet orbits in the upper panel, with arrows showing the acceleration of the star due to the disc $\mathbf{{a}_{*,d}} = -\ITd$ (starting from the star) and of the planet due to the direct force from the disc $\mathbf{{a}_{dir}}$ (starting at the planet location).
Note that in \myeqref{eq:acc_direct}, $\Sigma$ has been replaced by $\Sigma'$, as argued below, which does not change the $y$ component and the torque. 
The corresponding torque densities ($\Gamma' = d\Gamma/dr$) and integrated torque densities ($\int_0^r \Gamma'(x)dx$) are shown in the lower panel, normalized by $\Gamma_0$.
In this simulation, we find that $\Gamma_{\rm dir} = -1.00\,\Gamma_0$ while $$\Gamma_{\rm ind}=-0.08\,\Gamma_0\ = 0.08\,\Gamma_{\rm dir}\,.$$ 
The lower panel highlights that both $\Gamma_{\rm dir}$ and $\Gamma_{\rm ind}$ take most of their total value from the disc region very near the planet's orbital radius.
This is expected for the direct torque because this is where the wake is closest to the planet, and $\Gamma'_{\rm dir}$ is the highest. In contrast, $\Gamma'_{\rm ind}$ diverges when $r\to 0$; but its oscillations (due to the wrapping of the wake around the star) cancel out nicely. This shows that the choice for the radius of the inner edge of the simulation grid will not affect the results. The wrapping is distorted around the planet, and thus $\Gamma_{\rm ind}$ accumulates here. 
Also, the horseshoe region is asymetrical, with a surface density difference of $\sim0.01\Sigma_0$ between ahead and behind the planet, extending over $\sim 1$ radian; this leads to the direct corotation torque, but also contributes positively to $\mathbf{{a}_{*,d}}|_y$.

\begin{figure}
    \centering
    \includegraphics[width=\hsize]{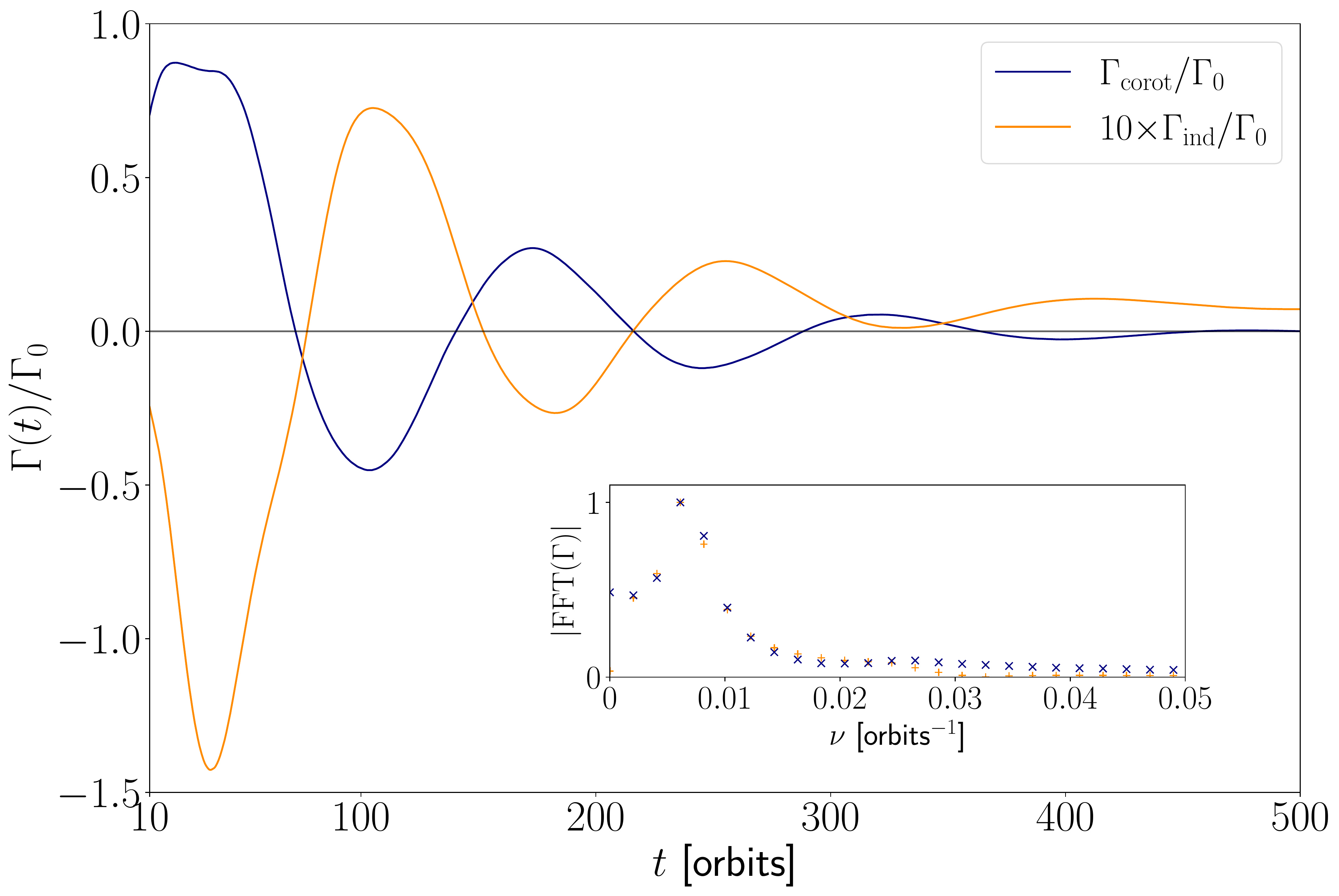}
    \caption{Time evolution of the corotation torque and the indirect torque in a simulation with a low-mass planet embedded in an inviscid disc (see text). Note that the indirect torque has been rescaled to improve legibility. The inset plot shows the normalised amplitude of the Fourier transforms of the corotation and indirect torques.}
    \label{fig:IT5b}
\end{figure}

To strengthen the link between the corotation torque and the indirect torque, we have run an additional simulation with the same parameters as above but with no turbulent viscosity, so that the corotation torque progressively cancels out. This is shown by the blue curve in Fig.~\ref{fig:IT5b}, where the corotation torque is calculated by simply subtracting from the direct torque its final value. The orange curve in the figure shows that the indirect torque also varies with time, and that it is actually anti-correlated with the corotation torque. This is further supported by the Fourier transforms of the time evolution of both torques, whose normalised amplitude is shown in the inset plot.

From these experiments we conclude that the indirect torque usually plays a quite modest role in type~I migration. Nonetheless, it can lead to a migration rate slightly different than what one would expect from $\Gamma_{\rm dir}$. Besides, $\Gamma_{\rm ind}/\Gamma_{\rm dir}$ depends on the physical parameters of the disc, including the slopes of the density and temperature profiles, as does the corotation torque.


\paragraph{Type I migration in presence of a vortex}
The indirect term can have a much higher impact if the disc features a strongly non-axisymmetric density distribution like that associated with a vortex, which results from instabilities setting in the disc.
The importance of the indirect torque in the interaction between a low-mass planet and a vortex has been put forward by \citet{Ataiee+2014}.
The vortex represents a mass excess similar to a secondary body, and the planet feels its gravitational pull much like a test particle would do in the restricted three-body problem, at least if the planet-to-vortex mass ratio remains small.
In particular, the vortex has its own $L_4$ and $L_5$ Lagrange points, where a planet can be captured as it migrates. 
This result may seem surprising at first since the direct torque from the disc is strictly positive (or negative) at these points. 
But the indirect term, which \citet{Ataiee+2014} call \emph{star torque} since it is the torque from the stellar gravity in the centre-of-mass frame, leads to a negative (resp. positive) torque, which balances exactly the direct term at $L_4$ and $L_5$. 
By discarding the indirect term from the disc, one would miss the capture of the planet at the Lagrange points of the vortex.

\begin{figure}
    \centering
    \includegraphics[width=\hsize]{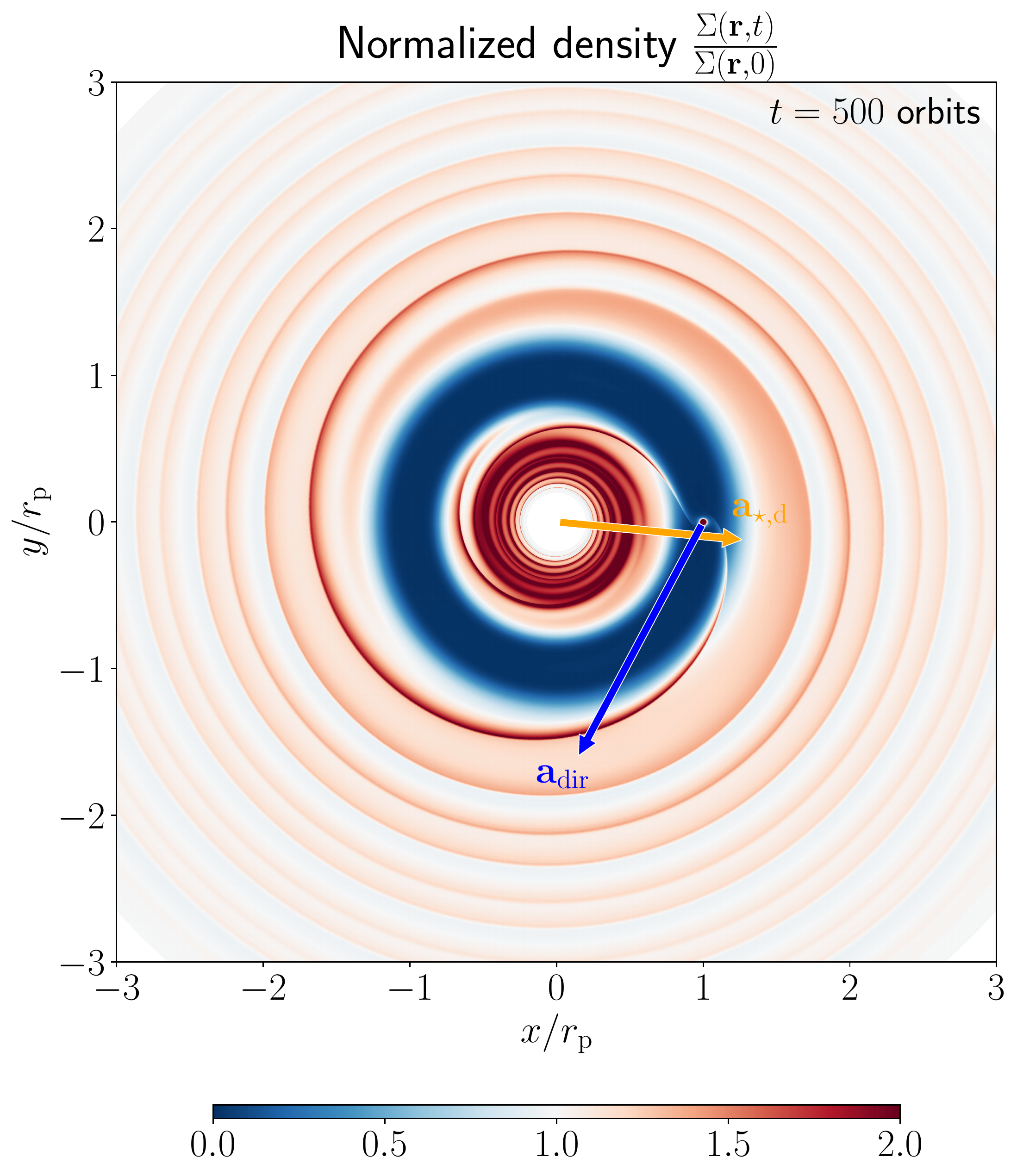}
    \includegraphics[width=\hsize]{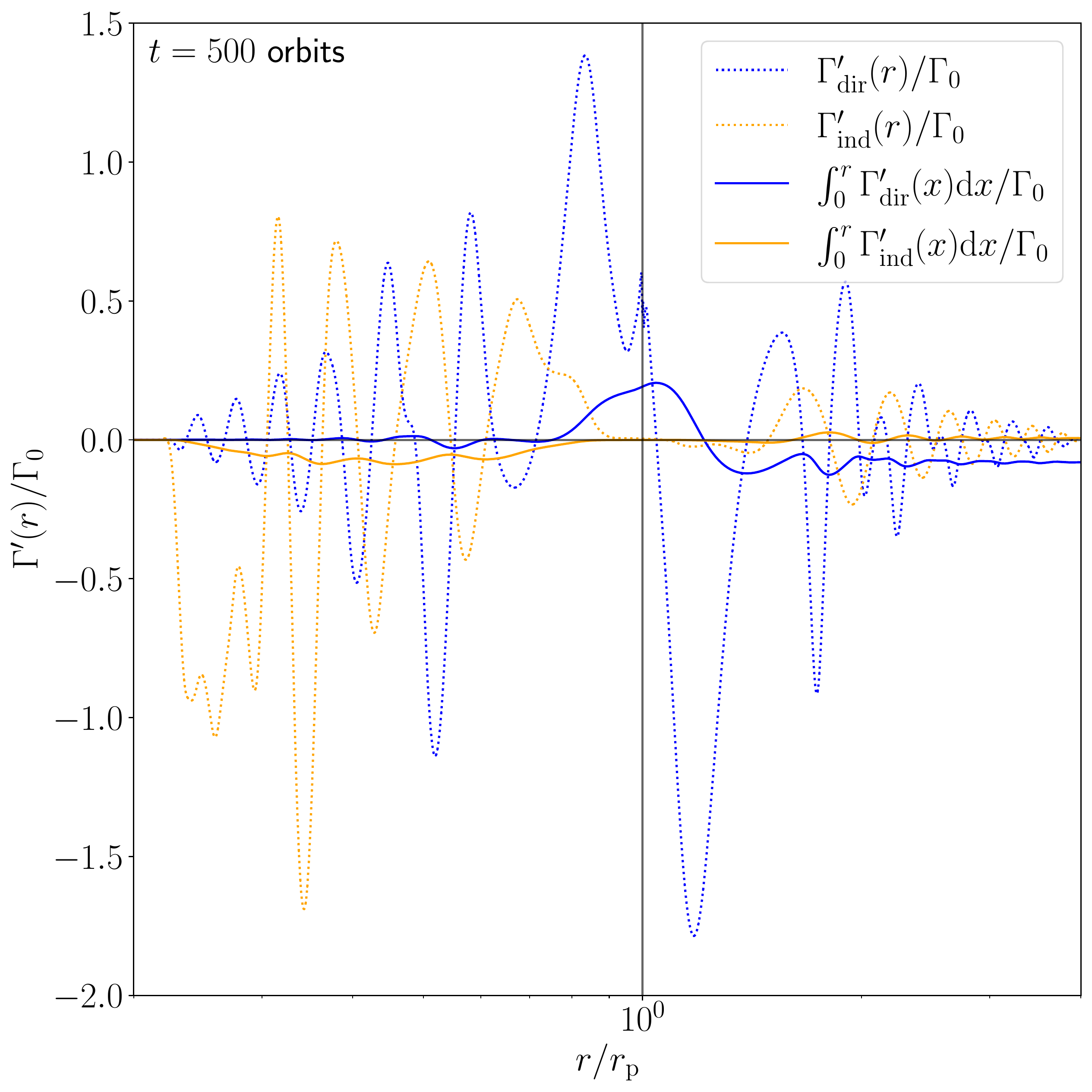}
    \caption{Same as \figref{fig:IT5} for a higher-mass planet that carves an annular gap around its orbit (see text). The scale for the arrows in the top panel is 1 length unit = $2\times 10^{-6}$ acceleration code unit.}
    \label{fig:IT3}
\end{figure}

\paragraph{Type~II migration}
As shown in Table~\ref{tab:simu}, we did another simulation similar to the one presented for the \textit{Type~I migration in a smooth disc} case (except for a slightly increased radius of the inner edge of the grid) with a planet-to-star mass ratio increased to $q=10^{-3}$, roughly equal to the mass ratio between Jupiter and the Sun.
\Figref{fig:IT3} displays again the disc gas density normalised by its initial profile, the torque densities and integrated torque densities.
Since the planet opens a substantial annular gap around its orbit, both the direct and indirect torques are reduced, but not in the same proportions. 
After 500 orbits, when the perturbed density has reached a near steady state, $\Gamma_{\rm dir} \sim -0.045\,\Gamma_0$ and $\Gamma_{\rm ind} \sim 0.0075\,\Gamma_0 \sim -0.16\,\Gamma_{\rm dir}$.
Still, during the first tens of orbits, the formation of vortices along the edges of the gap, as the latter gets progressively carved, results in a much larger indirect torque: $|\Gamma_{\rm ind}| \sim 0.6 |\Gamma_{\rm ind}|$. 
We see that, in this simulation, the indirect torque is definitely not negligible compared to the direct torque, and we stress that if the gap edges are not perfectly circular, this asymmetry will yield a strong indirect term \ITd.

\paragraph{Recipe for migrating planets}
The above paragraphs highlight that, in general, the indirect torque exerted by the disc on the planet should not be discarded, or, said differently, that both the direct and indirect torques of the disc should be computed to assess the planet's orbital evolution.
We conclude this subsection with a word of caution on the disc density to be used for calculating torques in disc-planet simulations.
From Eqs.~(\ref{eq:acc_direct}) to~(\ref{eq:Gamma_indirect}), it is clear that only the non-axisymmetric part of the disc's surface density can torque the planet, which is the quantity $\Sigma' = \Sigma - \langle\Sigma\rangle$ with $\langle\Sigma\rangle$ the azimuthally-averaged density.
$\langle\Sigma\rangle$ does exert a direct acceleration on the planet, but one that is oriented in the radial direction and which therefore does not torque the planet.
Still, when the disc's self-gravity is discarded, this radial acceleration introduces a spurious velocity difference between the planet and the background disc because only the planet feels the mass of the inner disc and thus orbits faster than around the star alone. This alters migration, as shown in \citet{PierensHure2005, BaruteauMasset2008SG, AtaieeKley2020}. 
A simple workaround for this issue in non self-gravitating discs simulations is to use $\Sigma'$ when computing the direct acceleration of the disc on the planet, as done for instance in the FARGO3D code\footnote{This requires adding the option \texttt{FARGO\_OPT += -DBM08} in FARGO3D's options setup. For the Dusty FARGO-ADSG code, this requires setting \texttt{BM08TRICK = Yes} in the parameter file.} \citep{BLM16}.
While the inclusion of $\langle\Sigma\rangle$ does alter the migration rate of planets, we have checked with a few dedicated simulations that it has little influence on the evolution of their eccentricity. In particular, the simulation shown in Fig.~\ref{fig:bp13} gives a very similar outcome whether one uses $\Sigma$ or $\Sigma'$ in \myeqref{eq:a_p_disc}.

In the end, we recommend that, in disc-planet simulations, the acceleration exerted by a non self-gravitating disc on a migrating planet should be
\begin{align}
    \mathbf{a_{{\rm disc}\rightarrow {\rm planet}}} 
   = 
    &\underbrace{\int_{\rm disc} \frac{G \Sigma'(\mathbf{r}) (\mathbf{r} - \mathbf{r_p})}{ \lVert \mathbf{r} - \mathbf{r_p} \rVert^3 } {\rm d}^2\mathbf{r}}_{\text{non-axisymmetric direct term}} \nonumber \\
   &- \underbrace{\int_{\rm disc} \frac{G \Sigma'(\mathbf{r}) \mathbf{r}}{ \lVert \mathbf{r} \rVert^3 } {\rm d}^2\mathbf{r}.}_{\text{indirect term}}
    \label{eq:a_p_disc}
\end{align}
In other words, only the non-axisymetric component of the disc density should be used.
We pledge for \myeqref{eq:a_p_disc} to be the new standard prescription for migrating planets in non self-gravitating discs, noting also that for massive planets, the integral should not be performed over the whole disc, but should exclude the circum-planetary disc \citep{Crida2009CPD}.

\ \\

\subsection{Indirect term from disc to disc (\ITdd)}
\label{sub:ITdd}

The indirect term of the disc is given by \myeqref{eq:ITd} above. 
It is imparted by the whole disc, as it is the sum of the indirect terms induced by each elementary disc element. The application of the collective $\mathbf{ITd}$ onto every disc element is noted \ITdd.
Whether or not \ITdd\ is implemented is most often \emph{not explicitly stated} in the literature. 
However, it may have important implications, as illustrated by the example cases in the next paragraphs.

\subsubsection{\ITdd\ and vortices}
Anticyclonic vortices are the natural manifestation of many instability mechanisms setting in gaseous protoplanetary discs.
They often merge to form a single, elongated vortex which imprints an indirect term on all constituents in the disc: planets (\ITdp, see Section~\ref{sub:ITdp}) and the disc itself (\ITdd, this paragraph).
 \citet{Zhu-Baruteau-2016} performed hydrodynamical simulations of a protoplanetary disc with an idealized density profile prone to the Rossby-Wave Instability (a Gaussian ring), in order to study the impact of the disc's self-gravity and indirect term on the growth and evolution of a vortex.
Their simulations showed that both the \emph{direct} (self-)gravitational acceleration of the vortex on the disc and its \emph{indirect} gravitational acceleration (\ITdd) readily impact the shape and evolution of the vortex for local Toomre Q parameters $\lesssim15$.
\footnote{\citet{Toomre1964}'s $Q$ parameter is the ratio of the dispersive (pressure and shear) forces to the self-gravity force. For a near Keplerian disc, $Q(R) \approx h(R) / \mu(R)$ with $h$ the disc's aspect ratio (ratio between the sound speed and the Keplerian velocity) and $\mu = [\pi\Sigma(R)R^2] / M_{\star}$ the reduced mass of the disc \citep[as introduced by][]{Crida-Morby-2007}, with $M_{\star}$ the mass of the star, and $\Sigma$ the disc's surface density. For a typical aspect ratio of h$=0.05$, $Q\lesssim 15$ corresponds to $\Sigma\gtrsim 10^{-3}\, M_\star R^{-2}$.}

When only \ITdd\ is accounted for, but not self-gravity, the vortex effectively excites an $m=1$ outer Lindblad resonance in the disc gas, which is very analogous to the case described in Section~\ref{sub:ITpp} of two planets entering their $2:1$ mean-motion resonance via their indirect term.
This $m=1$ outer Lindblad resonance takes the form of an outer spiral density wave (or wake) lagging the vortex, which causes the vortex to migrate inward towards the star and spin up.
The vortex thus becomes stronger, i.e. more compact and denser, as it spirals inward.
Now, when self-gravity is also taken into account, the vortex excites other Lindblad resonances in the disc, in particular inner Lindblad resonances that help couterbalance the sole effect of the $m=1$ outer Lindblad resonance brought about by \ITdd.
It implies that the vortex's inward migration is much reduced, and so is its spinning, as seen in \citet{Zhu-Baruteau-2016}. 
Their simulations with both self-gravity and \ITdd\ are closer to the ones with only self-gravity. 
Their simulations with none of these two terms are closer to the ones with only \ITdd, but with a weaker, non-migrating vortex. 

The analysis presented here allows us to conclude that including only the indirect term from the disc is the most incorrect choice.
It would correspond to having the disc evolve in the potential shown on the right panel of \figref{fig:RCTBP}, whose crescent shape can only enhance a vortex centred on $x=1, y=0$.
When self-gravity is discarded, the indirect term from the disc should not be applied to the disc (like for instance in \citealp{Robert+2020}) under penalty of having artificially strong, rapidly migrating vortices.
The \emph{vortex-driven migration} scenario of \citet{Lega+2021}, in which a vortex pushes a planet as the former migrates inward, is actually a consequence of fast vortex migration due to \ITdd, without disc self-gravity.
We have run new simulations with the same setup as in \citet{Lega+2021}, but without \ITdd, in which we do not recover this phase of vortex-driven planet migration.

\subsubsection{Influence on planetary migration}

\begin{table*}
    \centering
    \begin{tabular}{|l||c|c|c|}
    \hline
       ~ & planet 1 & planet 2 & disc\\
         \hline \hline
    planet 1\,: fixed &  $\mathbf{ITp_1p_1}$\,:\ \checkmark & $\mathbf{ITp_2p_1}$\,:\ $\times$  & \ITdp\,:\ $\times$\\
    \hline
    planet 1\,: migrating & $\mathbf{ITp_1p_1}$\,:\ \checkmark & $\mathbf{ITp_2p_1}$\,:\ $\square$  & \ITdp\,:\ \checkmark\\
     \hline
    disc element & \multicolumn{2}{c|}{\ITpd\,:\ \checkmark } & \ITdd\,:\ $\square^{\,*}$\\ 
    \hline
    \end{tabular}
    \caption{Table showing as columns the constituents of a protoplanetary disc from which an element in a row should feel the indirect term. Caption: \checkmark: yes, always. $\times$: no, never. $\square$: if and only if the direct term (e.g. self gravity for the disc) is also taken into account. $^*$ 
    See paper~II about the instability it can trigger.}
    \label{tab:tab}
\end{table*}

It should be noted that any response linear in gas surface density (for instance the response of a non self-gravitating disc to the gravitational perturbation by a small mass planet on a fixed circular orbit) generates an acceleration of the central star proportional to the disc mass. Hence, for different disc masses, \ITdd\ would yield different perturbed velocity and density fields, therefore
breaking the linearity of the response. One can have a strictly linear response of the disc only by working in a stellocentric frame and neglecting \ITdd.
This can be of importance for the study of type~I migration \citep[although the disc response is anyway not exactly linear as soon as the planet migrates; see figure~3 of][]{BaruteauMasset2008SG}.

In the case of giant planets, we have shown above that \ITdp\ plays a non negligible role, but \ITdd\ can be very important too, and not only for the artificial vortex-driven migration mentioned above. In numerical simulations of the migration of a pair of giant planets with \ITdd, \citet[][Chap. 5]{griveaud:tel-04831161} have observed episodes of outwards type~III migration similar to \citet[][figure 1]{Chametla+2020}. However, restarting the same simulation without \ITdd, these episodes disappear and the outwards migration of our Jupiter - Saturn pair is smooth. Our interpretation is that \ITdd\ excites some perturbations of the density in the disc (see below) which allow the coorbital mass deficit to temporarily exceed the outer planet mass and trigger a runaway migration episode \citep{Masset-Papaloizou-2003}. It highlights that explicitly saying what is done with \ITdd\ is necessary for the reproducibility of a study.

\subsubsection{Does \ITdd\ make discs unstable?}

As soon as a protoplanetary disc becomes non-axisymmetric, be it by the onset of instabilities or protoplanet formation, \ITdd\ becomes non-zero.
\ITdd\ perturbs the disc in return, 
which can reinforce \ITdd, thus providing a positive feedback loop. 
In Paper~II, we show that \ITdd\ grows an instability associated with an $m=1$ mode (in other words, it makes the disk eccentric), even when the disc's self-gravity is included.
It is found in 2D and 3D hydrodynamical simulations, with finite-difference and volume-finite grid-based codes, and with an SPH code.
It is also found in simulations that solve the governing equations in an inertial frame of reference centred on the centre-of-mass of the system.
All this indicates that the instability is of physical origin, and not due to numerical artefacts associated with boundary conditions or the implementation of the disc's indirect term.
More details will be found in Paper~II, where we show how the growth timescale of the instability depends on the disc's total mass. 
For typical low disc-to-star mass ratios of order a percent or so, the instability can take a few thousand dynamical timescales, if not more, to manifest itself with large amplitude perturbations in the disc. 
But before destabilising the disc, the $m=1$ mode can lead to a spurious oscillating torque felt by a planet, a troubling phenomenon that we have already encountered without understanding its nature. 
We found that removing \ITdd\ suppresses this effect.

\section{Conclusion}

In this article, we have shown that when gravitational systems are described in a frame centred on the most massive, primary object, there is not \emph{an} indirect term but \emph{multiple} indirect terms, which are as many as there are constituents that exert a gravitational pull on the primary (e.g. planets, gaseous discs).
When all the gravitational interactions are modelled, in particular the gas self-gravity, all components of the indirect term should be taken into account without question. But when the direct gravitational force from one constituent in the system is discarded, so should be its indirect term.
Conversely, when the direct gravitational force from one constituent is included, so should be its corresponding indirect term.
These ideas are summarised in table~\ref{tab:tab} for the specific case of a system comprised of a star, a protoplanetary disc and planets.
We advise all readers to follow this table, but in any case one should mention explicitly if \ITdp\ and \ITdd\ are included in their simulations. Most often, numerical simulations only mention \emph{one} indirect term, so that $\mathbf{ITd}$ is applied to all or none of the objects. 
We stress here that it must be applied to migrating planets, but that it is better to not apply it to a non-self-gravitating gas disc. Hence, indirect terms should be separated and each individual indirect term should be judiciously included or otherwise.

We have also provided a recipe for the force to be applied on a migrating planet embedded in a non-self-gravitating discs, which is given by \myeqref{eq:a_p_disc}. 
Almost all simulations of planet migration in non-self-gravitating discs published so far are inconsistent for the following reasons: either they do not include the indirect term of the disc while including the direct gravity of the disc on planets, or they do include the disc indirect term but not the direct gravity of the disc on itself (self-gravity).
Fortunately, the effects of \ITdd\ often appear over long times, especially in less massive discs, so most results published in the literature so far are just fine. 
However, in low-viscosity discs featuring vortices and/or in the most massive discs, \ITdd\ can perturb significantly and unexpectedly the disc dynamics (see Paper~II), and thus should be handled carefully.

\vfill
\section*{Acknowledgements}
We would like to warmly thank the "Programme National de Physique Stellaire" (PNPS) and "Programme National de Planétologie" (PNP) of CNRS/INSU co-funded by CEA and CNES for their financial support of our BiPhasique research group. FM acknowledges support from UNAM's grant PAPIIT 107723, UNAM's DGAPA PASPA program and the Laboratoire Lagrange at Observatoire de la C\^ote d'Azur for hospitality during a one-year sabbatical stay. JFG thanks the LABEX Lyon Institute of Origins (ANR-10-LABX-0066) for its financial support within the Plan France 2030 of the French government operated by the ANR. This work was performed using HPC resources Mesocentre SIGAMM, hosted by the Observatoire the la C\^ote d'Azur. 

\newpage
\begin{appendix}

\section{A) The indirect term in the two-body problem}
\label{sub:appA}
Let us consider the two-body problem with a planet of mass $M_p$ on a circular orbit of radius $r_p$ and angular frequency $\Omega$ around a star of mass $M_*$. 
Noting $M=M_*+M_p$ and $\mu=M_p/M$, the centre-of-mass of the system, $C$, is located at a distance $\mu r_p$ from the star in the direction of the planet. 
In a frame centred on C and rotating at angular frequency $\Omega$, the planet and the star are stationary. 
The planet feels an acceleration $GM_*/{r_p}^2$ in the direction of $C$ from the star, and a centrifugal acceleration $[(1-\mu)r_p]\Omega^2=(M_*/M)r_p\Omega^2$ in the direction away from $C$. 
They balance each other exactly, leading to the famous $\Omega^2 = GM/{r_p}^3$, where again $M=M_*+M_p$.

The same analysis performed in a frame centred on the star would yield a centrifugal acceleration $r_p\Omega^2$, and in the end $\Omega^2 = GM_*/{r_p}^3 < GM/{r_p}^3$. 
This is wrong, and can be corrected by accounting for the indirect term due to the planet, given in \myeqref{eq:ITp}.
With this indirect term, the balance of accelerations on the planet, projected on the x-axis, reads: 
\begin{equation}
    0 = -\frac{GM_*}{{r_p}^2} + {r_p}\Omega^2 -\frac{GM_p}{{r_p}^2},
\end{equation}
which yields again $\Omega^2 = GM / {r_p}^3$.
In other words, the indirect term is responsible for the planet to rotate around the star at the angular frequency $\sqrt{G(M_*+M_p)/{r_p}^3}$ instead of $\sqrt{GM_*/{r_p}^3}$. 

\ \\
\section{B) Resonances due to the Indirect Term}
\label{sub:appB}

In this appendix, we demonstrate how the indirect term can excite or capture elements with the 2:1 outer Lindblad resonance, but with no other. The dynamics associated with $\mathbf{IT}$ is therefore very peculiar and can lead to surprises, misunderstandings, and spurious phenomena if it is not removed properly when the direct term itself is turned off. Only with both the direct and the indirect terms are Lindblad resonances properly modelled. For the interest of the readers, this is shown using a classical and a Hamiltonian approach.

\paragraph{Classical physics}
As shown in \myeqref{eq:a_2}, the influence of a body of mass $M_1$ orbiting around the star with a position vector $\mathbf{r_1}$ (assumed of constant norm) on another body located at $\mathbf{r_2}$ can be split into an acceleration along $\mathbf{r_2}$ and another along $\mathbf{r_1}$. The latter is the one which can torque object 2 and reads\,:
$$
   \mathbf{a_2|_{r_1}} = - \frac{GM_1}{{r_1}^3}\mathbf{r_1} + \frac{GM_1}{d^3}\mathbf{r_1},
$$
where $d=|\mathbf{r_2}-\mathbf{r_1}|$. The first term is the indirect term and the second term is a component of the direct acceleration from $M_1$, $- \frac{GM_1}{{|\mathbf{r_1}-\mathbf{r_2}|}^3}(\mathbf{r_1}-\mathbf{r_2})$. Noting $\theta_i$ the azimuth (true longitude) of body $i$, these two terms are $2\pi$-periodic functions of $\vartheta=\theta_1-\theta_2$ and we can decompose them in Fourier series\,:
\begin{eqnarray}
    \mathbf{r_1} & = & r_1\cdot\left(\cos(\vartheta)\mathbf{\hat{r_2}} +\sin(\vartheta)\mathbf{\hat{\theta_2}}\right)\\
    \displaystyle\frac{1}{d} & = & \frac{1}{r_1} \sum_{m=0}^\infty b_{1/2}^{(m)} \left(\frac{r_2}{r_1}\right)\cos(m\vartheta)
\end{eqnarray}
where $(\mathbf{\hat{r}_2},\mathbf{\hat{\theta}_2})$ is the base of unit vectors in polar coordinates centred on the star and attached to object 2, and $b_p^{(m)}(\alpha)$ are the Laplace coefficients. We see that the first term in $\mathbf{a_2|_{r_1}}$ only has a Fourier component with azimuthal wavenumber $m=1$. It is of constant amplitude and its direction rotates periodically (with the period of $\vartheta$, that is the synodic period from the point of view of body 2). 
In contrast, because it depends on $1/d$, the second term of $\mathbf{a_2|_{r_1}}$ is a combination of modes of all $m\in\mathbb{N}$. As a consequence, the various harmonics of the direct term from $M_1$ can excite resonances with the proper epicyclic frequency of body 2, $\kappa_2$, leading to all resonance orders\,: there is a Lindblad resonance of order $n$ whenever $\kappa_2=m\dot\vartheta$ (where $\dot\vartheta=\Omega_1-\Omega_2$ is the time derivative of $\vartheta$), that is --\,assuming $\kappa_2\approx\Omega_2$\,-- when $(m+1)\Omega_2 = m\Omega_1$. 

But the indirect term can only resonate with the horizontal epicyclic motion when $\kappa_2=\dot\vartheta$, which corresponds to the outer 2:1 Lindblad resonance with body 1 ($2\Omega_2 = \Omega_1$).

\paragraph{Hamiltonian formalism}
A similar argument can be made within the Hamiltonian formalism. 
In this case, the reflex motion of the star around the centre of mass of the system appears in the kinetic energy of the system. With $M_*\mathbf{v}_* = -M_1\dot{\mathbf{r}}_1-M_2\dot{\mathbf{r}}_2$, the kinetic energy of the star $\frac12M_*\mathbf{v_*}^2$ has only one term that involves both planets and thus can describe the influence they indirectly have on each other \citep[Eq.~(8) of][]{Laskar-Robutel-1995}\,:
\begin{equation*}
\mathcal{T}_1 = \frac{M_1 M_2}{M_*} \dot{\mathbf{r}}_1 \cdotp \dot{\mathbf{r}}_2 = \frac{M_1 M_2}{M_*} \left(\dot{x}_1 \dot{x}_2 + \dot{y}_1 \dot{y}_2\right).
\end{equation*}
In the orbital plane with a reference frame aligned with the pericentre, the velocity of a planet is given in the two-body problem by \citep[see e.g.][sect.~5]{Laskar-Boue-2010}\,:
$$\dot{\mathbf{r}} = \left\{
\begin{array}{rl}
\dot{x} &= - \Omega a\left(\frac{a}{r} \sin{E}\right),\vspace{3pt}\\
\dot{y} &= \Omega a \sqrt{1-e^2}\left(\frac{a}{r} \cos{E}\right)
\end{array}
\right.$$
where $E$ is the eccentric anomaly. The terms in parenthesis are then expanded classically as
\begin{align*}
\frac{a}{r} \cos{E} &= \frac{2}{e} \sum_{k=1}^{\infty} J_k (ke) \cos{k \mathcal{M}} = \cos\mathcal{M} + e\cos(2\mathcal{M}) + \mathcal{O}(e^2)\\
\frac{a}{r} \sin{E} &= 2 \sum_{k=1}^{\infty} J'_k (ke) \sin{k \mathcal{M}} = \sin\mathcal{M} + e\sin(2\mathcal{M}) + \mathcal{O}(e^2)
\end{align*}
where $\mathcal{M}$ is the mean anomaly and $J_k$ are the Bessel functions. Since $J_1(e)\propto e$ and $J_2(2e)\propto e^2$, at lowest order in $e$ the only combinations of angles $\mathcal{M}_1$, $\mathcal{M}_2$ that can appear in $\mathcal{T}_1$ are $\mathcal{M}_1-\mathcal{M}_2$ (independent of eccentricities), $\mathcal{M}_1 - 2\mathcal{M}_2$ (proportional to $e_2$), and $2\mathcal{M}_1 - \mathcal{M}_2$ (proportional to $e_1$). The latter two are the contribution of the indirect term to the 2:1 outer Lindblad resonance with bodies 1 and 2 respectively\,: their time derivative cancel (so they don't average to zero on secular timescales) when $\Omega_1=2\Omega_2$ and $2\Omega_1=\Omega_2$ respectively.

\end{appendix}


\end{document}